\documentclass[11pt]{article}

\usepackage{url}
\usepackage{graphicx}
\usepackage{amsmath}
\usepackage{amssymb}
\usepackage{color}

\usepackage{makeidx}

\makeindex

\allowdisplaybreaks[2]

\newcommand{\gev}{\operatorname{GeV}}
\newcommand{\mev}{\operatorname{MeV}}

\newcommand{\ud}{\mathrm{d}}

\newcommand{\gsim}{\raisebox{-4pt}{%
    $\,\stackrel{\textstyle >}{\sim}\,$}}

\begin{document}

{\huge {\bf  The gluon Sivers distribution:\\[1ex]
\mbox{}\hspace{\parindent}status and future prospects}\\[1 ex]}

\parbox{6in}{\slshape
Dani\"el Boer$\,^1$, C\'edric Lorc\'e$\,^{2,3}$, Cristian Pisano$\,^4$, Jian Zhou$\,^5$\\[1ex]
$^1$\ Van Swinderen Institute, University of Groningen, Nijenborgh 4, 9747 AG Groningen, The Netherlands  \\
$^2$\  SLAC National Accelerator Laboratory, Stanford University, Menlo Park, CA 94025, USA \\
$^3$\  IFPA,  AGO Department, Universit\'e de Li\` ege, Sart-Tilman, 4000 Li\`ege, Belgium \\
$^4$\ Department of Physics, University of Antwerp, Groenenborgerlaan 171, 2020 Antwerp, Belgium\\
$^5$\ Nikhef and Department of Physics and Astronomy, VU University Amsterdam,  De Boelelaan 1081, NL-1081 HV Amsterdam, The Netherlands}\\[1ex]

\centerline{\bf Abstract}
\vspace{0.7\baselineskip}
\parbox{0.9\textwidth}{}
\noindent
This is a review of what is currently known about the gluon Sivers distribution and of what are the opportunities to learn more about it.
Because single transverse spin asymmetries in $p^\uparrow \, p \to \pi \, X$ provide only indirect information about the gluon Sivers function through the relation with the quark-gluon and tri-gluon Qiu-Sterman functions, current data from hadronic collisions at RHIC have not yet been translated into a solid constraint on the gluon Sivers function. SIDIS data, including the COMPASS deuteron data, allow for a gluon Sivers contribution that is of the natural size expected from large $N_c$ arguments, which is ${\cal O}(1/N_c)$ times the nonsinglet quark Sivers contribution. Several very promising processes to measure the gluon Sivers effect directly have been put forward, which apart from ongoing and future investigations at RHIC, would strongly favor experiments at AFTER@LHC and a possible future Electron-Ion Collider. Due to the inherent process dependence of TMDs, the gluon Sivers TMD probed in the various processes are not necessarily the same, but rather can be different linear combinations of two universal gluon Sivers functions that have different behavior under charge conjugation and that therefore satisfy different theoretical constraints. For this reason both hadronic and DIS type of collisions are essential in the study of the role of gluons in transversely polarized protons.


\section{The Sivers function and its definition}

The distribution of quarks and gluons in a proton (or any other spin-$1/2$ hadron) that is polarized transversely to its momentum need not be left-right symmetric w.r.t.\ the plane spanned by the momentum and spin directions. This asymmetry is called the Sivers effect~\cite{Sivers:1989cc}. It results in angular asymmetries of produced particles in high energy scattering processes involving a transversely polarized hadron. Experimental data in support for such a left-right asymmetry in the quark distribution was first obtained from semi-inclusive DIS process by the HERMES collaboration~\cite{Airapetian:2004tw}. This review is about what is currently known about the gluonic Sivers effect distribution. The Sivers effect is of great interest theoretically as it is very sensitive to the color flow in the scattering process and to the multitude of color exchanges among initial and final states. It is the first quantity for which this has been recognized and for which color flow sensitivity can be tested unambiguously~\cite{Collins:2002kn}. Verification of its unusual properties will provide a strong test of the formalism of transverse momentum dependent parton distributions. It is an important quantity of nonperturbative QCD to consider both qualitatively and quantitatively. 
This review discusses these aspects for the gluon Sivers distribution specifically. We first start with its proper definition.

The number density in momentum space of a generic parton  (quark, antiquark, gluon) inside a hadron with mass $M$, transverse
polarization $\boldsymbol{S}_{\scriptscriptstyle T}$ and momentum $\boldsymbol{P}$ can be written as
\begin{equation}
\hat {f}  (x,\boldsymbol k_\perp;
\boldsymbol S_{\scriptscriptstyle T}) =
f_1 (x,\boldsymbol k_\perp^2)
-\frac{(\hat{\boldsymbol P}\times \boldsymbol k_\perp )\cdot \hat{\boldsymbol S}_{\scriptscriptstyle T}}{M}\, f^{\perp}_{1 T}
(x,\boldsymbol k_\perp^2) \, ,
\label{eq:def}
\end{equation}
where $f_1 (x,\boldsymbol k_\perp^2)$ is the unpolarized Transverse-Momentum Dependent (TMD) parton distribution,
$\hat{\boldsymbol S}_{\scriptscriptstyle T} \equiv \boldsymbol S_{\scriptscriptstyle T} /\vert \boldsymbol S_{\scriptscriptstyle T}\vert$, and $\hat{\boldsymbol P} \equiv\boldsymbol P/ \vert \boldsymbol P\vert$.
The function $f^{\perp}_{1 T}(x, \boldsymbol k_\perp^2)$ describing the distortion in the distribution of unpolarized partons with light-front momentum fraction $x$ and transverse momentum $\boldsymbol k_{\perp}$ due to the transverse polarization of the hadron is called the Sivers function. The notation used here comes from~\cite{Boer:1997nt}, but also the notation
$\Delta^N f_{g/h^{\uparrow}}$ from~\cite{Anselmino:1994tv,Anselmino:2005sh} is sometimes used, where $ \Delta^N f_{g/h^{\uparrow}} = -2\, \frac{\vert\boldsymbol k_\perp\vert}{M}\, f^{\perp\, g}_{1 T}$ (analogous to the quark case~\cite{Bacchetta:2004jz}).

The Sivers function satisfies the following positivity bound~\cite{Mulders:2000sh}:
\begin{equation}
\frac {\vert \boldsymbol k_\perp\vert}{M}\,  \vert f^{\perp}_{1T} (x,  \boldsymbol k_\perp^2) \vert \le
 f_1 (x, \boldsymbol k_\perp^2 ).
\end{equation}
In Ref.~\cite{Mulders:2000sh} the operator definition of the gluon Sivers function, which was called $G_{T} = -f^{\perp\, g}_{1 T}$~\cite{Meissner:2007rx}, was first given without gauge links. The definition including gauge links then appeared in Refs.~\cite{Lorce:2013pza,Buffing:2013kca}
\begin{equation}
-\frac{(\hat{\boldsymbol P}\times \boldsymbol k_\perp )\cdot \hat{\boldsymbol S}_{\scriptscriptstyle T}}{M}\,f^{\perp g}_{1T}(x,\boldsymbol k_\perp^2)=\frac{1}{2}\left[\hat {f}  (x,\boldsymbol k_\perp;
\boldsymbol S_{\scriptscriptstyle T})-\hat {f}  (x,\boldsymbol k_\perp;
-\boldsymbol S_{\scriptscriptstyle T}) \right]
\end{equation}
with
\begin{equation}
\hat {f}  (x,\boldsymbol k_\perp;
\boldsymbol S_{\scriptscriptstyle T})=\frac{\delta^{jl}_{T}}{xP^+}\int\frac{\ud z^-\,\ud^2z_\perp}{(2\pi)^2}\,e^{ik\cdot z}\,\langle P,S_T|\,2\mathrm{Tr}\!\left[F^{+j}(0)U_{[0,z]}F^{+l}(z)U_{[z,0]}\right]|P,S_T\rangle\Big|_{z^+=0},
\end{equation}
where $U_{[a,b]}$ is a Wilson line connecting the points $a$ and $b$ along a contour determined by the physical process and $k^+=xP^+$ is the fraction of parton light-front momentum. For a proper definition that is free from rapidity divergences associated with gauge links with paths (partly) along the light front, a redefinition involving the so-called soft factor is necessary~\cite{Collins:2011zzd,Collins:2011ca,Echevarria:2012js,Echevarria:2015uaa}. This will however not play a significant role here and can simply be considered as implicit.

\section{Sivers effect, $A_N$, and Qiu-Sterman effect}

The Sivers effect (for both quarks and gluons) was first suggested in~\cite{Sivers:1989cc} as an explanation for the large left-right single transverse spin asymmetries ($A_N$) observed in $p^\uparrow \, p \to \pi \, X$ \cite{Antille:1980th,Adams:1991cs,Adams:1991rw,Krueger:1998hz,Adams:2003fx,Adler:2005in,Arsene:2008aa,Abelev:2008af} (and similar asymmetries in $K$ \cite{Arsene:2008aa}, in $\eta$ \cite{Adamczyk:2012xd}, and tentatively in $J/\psi$ \cite{Adare:2010bd} production). The Sivers effect was first studied phenomenologically in~\cite{Anselmino:1994tv}. Extraction of the Sivers TMD presumes all-order TMD factorization, however. A factorized description of the process $p \, p \to \pi \, X$ only applies for large transverse momentum $p_T$ of the produced pions (say for $p_T \gsim 1\ {\rm GeV}$), where in fact collinear factorization is appropriate~\cite{Collins:1989gx}, rather than TMD factorization. Although TMDs do appear in the phenomenological description of $A_N$ in \cite{Anselmino:1994tv} and subsequent studies (see \cite{Anselmino:2015eoa}), that description is thus not based on a TMD factorization theorem. Rather it should be considered as an effective model description, now commonly referred to as the Generalized Parton Model (GPM). As a phenomenological approach it has proven useful in the quest to disentangle the possible underlying mechanisms of the spin asymmetries, see~\cite{Anselmino:2015eoa,Kanazawa:2015fia} for more discussion, but the extracted ``effective'' TMDs may differ from the TMDs extracted from TMD factorizing processes. What is known about the effective gluon Sivers TMD will be discussed below. 

In collinear factorization the single spin asymmetry (SSA) will arise at the twist-3 level~\cite{Luo:1991bj}. In this description $A_N$ probes the (quark-gluon) Qiu-Sterman function\footnote{Sometimes one factor of the coupling constant $g$ is included in the definition of Qiu-Sterman functions, because one always encounters them multiplied by $g$.}~\cite{Qiu:1991pp,Qiu:1998ia}
\begin{equation}
\begin{split}
T_{q,F}&(x,x)=\\
&\frac{M}{P^+}\int\frac{\ud z^-\,\ud\eta}{2\pi}\,e^{ik\cdot z}\,\frac{(\hat{\boldsymbol P}\times\hat{\boldsymbol S}_T)^j}{2M}\,\langle P,S|\overline\psi(0)\gamma^+F^{+j}(\eta z)\psi(z)|P,S\rangle\Big|_{z^+=\vert \boldsymbol z_\perp\vert=0},
\end{split}
\end{equation}
and its tri-gluon correlation analogues $T_G^{(f)}(x,x)=T_G^{(+)}(x,x)$ and $T_G^{(d)}(x,x)=T_G^{(-)}(x,x)$ \cite{Ji:1992eu,Kang:2008qh,Kang:2008ih,Schafer:2013opa}
\begin{equation}
\begin{split}
T_{G}^{(\pm)}&(x,x)=\\
&-\frac{2M\delta^{lm}_{T}}{x(P^+)^2}\int\frac{\ud z^-\,\ud\eta}{2\pi}\,e^{ik\cdot z}\,\frac{(\hat{\boldsymbol P}\times\hat{\boldsymbol S}_T)^j}{2M}\,\langle P,S|C^{abc}_\pm F_a^{+l}(0)F_b^{+j}(\eta z)F_c^{+m}(z)|P,S\rangle\Big|_{z^+=\vert \boldsymbol z_\perp\vert=0},
\end{split}
\end{equation}
where $C_+^{abc}=if^{abc}$ and $C_-^{abc}=d^{abc}$, and where the light-front gauge $A^+=0$ has been considered for convenience.

In~\cite{Boer:2003cm} the quark-gluon Qiu-Sterman function has been related to the first transverse moment of the quark Sivers function, i.e.\
$f^{\perp(1)q}_{1T}(x) \propto T_{q,F}(x,x)/M$, where
\begin{equation}\label{transvmom}
f^{\perp(1)q}_{1T}(x)\equiv\int\ud^2k_\perp\,\frac{\boldsymbol k^2_\perp}{2M^2}\,f^{\perp q}_{1T}(x,\boldsymbol k^2_\perp).
\end{equation}
However, that relation was only established at tree level (beyond tree level the relation will be affected by the considered regularized definition of the Sivers TMD including its dependence on the soft factor). A similar tree level relation can be established in the gluon sector as well: $f^{\perp(1)g}_{1T}(x) \propto T_{G}(x,x)/M$ (which $T_G$ appears in this relation depends on the gauge links, see the discussion in section \ref{sect:procdep}).

Another relation has been established in~\cite{Ji:2006ub}:
\begin{equation}
f_{1T}^{\perp q}(x,\boldsymbol{k}_\perp^2) \stackrel{\boldsymbol{k}_\perp^2 \gg M^2}{\sim} \alpha_s\, \frac{M}{\boldsymbol{k}_\perp^4} \, \left(K \otimes T_{q,F}\right) (x), \label{tail}
\end{equation}
which means that the quark-gluon Qiu-Sterman function determines the large transverse momentum tail of the quark Sivers function. Here it should be emphasized that the function $\left(K \otimes T_{q,F}\right) (x)$ consists not only of a convolution of $T_{q,F}(x,x)$, but also of its derivative $x \partial T_{q,F}(x,x)/\partial x$ and of the more general $T_{q,F}(x,y)$ with $y\neq x$. It corresponds to the fact that the evolution of $T_{q,F}$ is non-autonomous and inhomogeneous, see \cite{Braun:2009mi}. Note that here we have discussed the non-singlet contributions only, that apply to combinations like $u$ minus $d$ quarks, otherwise also gluonic contributions need to be taken into account.

Similarly, the tail of the gluon Sivers function is determined by several Qiu-Sterman functions~\cite{Schafer:2013wca}. It receives contributions from the quark-gluon Qiu-Sterman functions $T_{q,F}(x,x)$ and $T_{q,F}(x,y)$ with $y\neq x$, and from the tri-gluon functions $T_G^{(f/d)}(x,x)$ (which one(s) depends on the gauge links, see section \ref{sect:procdep}). At small $x$ the situation simplifies:
the contributions from $T_{q,F}(x,x)$ and $T_{q,F}(x,y)$ with $y\neq x$ to the tail of the gluon Sivers function cancel each other~\cite{Schafer:2013wca}, leaving only the tri-gluon correlators. Moreover, $T_G^{(d)}(x,x)$ evolves with the same $1/x$ behavior at small $x$ as the unpolarized gluon distribution, and is therefore not necessarily suppressed at high energies and small values of $x$, whereas $T_G^{(f)}(x,x)$ lacks this $1/x$ enhancement~\cite{Schafer:2013opa}. As the large $p_T$ $A_N$ data from RHIC are generally not in the small-$x$ region of the polarized proton, except for negative $x_F$, any simplifications at small $x$ should of course first be tested for validity.

Information from $A_N$ measurements at sufficiently large $p_T$ (in order to consider a collinear factorization description in the first place) can thus {\it in principle} reflect some information on Sivers functions (i.e.\ on the tails and perhaps also on first transverse moments), but {\it in practice} other twist-3 contributions beside the mentioned Qiu-Sterman functions, namely chiral-odd and fragmentation function analogues, contribute to $A_N$~\cite{Kanazawa:2010au,Kanazawa:2011bg,Kanazawa:2014dca,Kanazawa:2015fia}. From the smallness of $A_N$ in the midrapity and backward (negative $x_F$) regions, one would generally conclude  that gluonic and sea quark contributions to the transverse single spin asymmetry are not large, but a detailed analysis is required to determine precisely the size of the various contributions. 

$A_N$ for $\pi^0$ production at midrapidity has been measured by the PHENIX experiment in polarized $p\,p$ collisions at RHIC and was found to be consistent with zero, for $p_T$ values below 5 GeV at the permille level and for higher $p_T$ values (up to 11 GeV) at the few percent level~\cite{Adler:2005in,Adare:2013ekj}. These data taken at $\sqrt{s}=200$ GeV probe $x$ values only down to $x \sim 0.006$, where still a combination of Qiu-Sterman functions is expected to contribute. In \cite{Beppu:2013uda} these $\pi^0$ data were discussed, using two models for the tri-gluon Qiu-Sterman functions that were constrained from $A_N$ in $D$-meson production~\cite{Liu:2009zzw,Koike:2011mb}. The midrapidity $\pi^0$ data are shown to mostly constrain $T_G^{(f)}(x,x)$ (their $N(x)$) at low $p_T$. The authors conclude that ``Both models give tiny asymmetry due to the small partonic cross sections, so the form of the three-gluon correlation functions is not much constrained by the data in this region.'' From this limited model study of both $\pi^0$ and $D$ production one would conclude that $T_G^{(f)}(x,x)$ and $T_G^{(d)}(x,x)$ are in any case small, a permille fraction of $x$ times the unpolarized gluon distribution. This should be investigated further with more general model forms that adhere to the correct small-$x$ behavior and with more precise data. The experimental precision of $A_N$ can be improved much further both at RHIC and especially at the AFTER@LHC experiment~\cite{Brodsky:2012vg}, which would have a luminosity factor of 10-100 higher, if not more. Such improvement and the measurements of asymmetries for many different types of produced particles are required to separately constrain or determine the various Qiu-Sterman contributions. 

In the GPM the smallness of $A_N$ at midrapidity puts strong constraints on the effective gluon Sivers function. As explained, this gluon Sivers function captures the combined effect from several Qiu-Sterman contributions and may thus differ from the actual gluon Sivers function obtained from TMD-factorizing processes. In a recent GPM analysis~\cite{D'Alesio:2015uta}, which is an updated analysis of Ref.~\cite{Anselmino:2006yq}, the {\it best fits} to the PHENIX midrapidity $\pi^0$ $A_N$ data indeed correspond to a small effective gluon Sivers function w.r.t.\ its theoretical bound determined by the unpolarized gluon. For example, for $x <0.1$ it is at most only a few percent of the bound. However, the maximally allowed effective gluon Sivers function is still sizable though. Its first transverse moment is still found to be around 30\% of the up quark Sivers function in the region $0.06 < x < 0.3$, which is consistent with findings from semi-inclusive DIS for the actual gluon Sivers function and also with theoretical expectations, as discussed in the next section. In addition, it should be mentioned that this GPM analysis assumes a Gaussian $k_\perp$ dependence, which does not correspond to the correct power-law tail of the Sivers function, Eq.\ (\ref{tail}), nor of the unpolarized gluon distribution. Given all the caveats that come with these results, one should be careful to draw a definite conclusion about the size of the actual gluon Sivers TMD from $A_N$ data. 

One should also specify clearly what one calls a small gluon Sivers function. It will depend on what one compares to, i.e.\ whether that is to the unpolarized gluon that grows very rapidly at small $x$, or to the up or down quark Sivers function for not too small $x$. 
At small $x$ it becomes very important whether one discusses the $f$ or $d$ type contribution (see section \ref{sect:procdep}), which is an issue not addressed in the GPM studies of $A_N$.

\section{Sivers asymmetry in SIDIS}

The Sivers effect leads to a $\sin(\phi_h-\phi_S)$ asymmetry in semi-inclusive DIS (SIDIS)~\cite{Boer:1997nt}, which has been observed in experiments using a proton target by HERMES~\cite{Airapetian:2004tw,Airapetian:2009ae} and COMPASS~\cite{Adolph:2012sp}, and using a $^3$He target by Jefferson Lab Hall A~\cite{Qian:2011py,Zhao:2014qvx}. The data follow to quite a good extent the expectations of a valence quark picture in the target and of favored fragmentation. In the proton case, the $\pi^+$ thus shows the largest asymmetry (for large $z$ values, the asymmetry is around 4-5\%, or even somewhat larger when a lower cut of $Q^2>4\ {\rm GeV}^2$ is implemented instead of $Q^2>1\ {\rm GeV}^2$~\cite{Airapetian:2009ae}). The $\pi^-$ asymmetries are smaller and still compatible with zero. The $K^\pm$ asymmetries are similar to the $\pi^{\pm}$ asymmetries, but with larger errors. Sivers asymmetries on a deuteron target~\cite{Alekseev:2008aa} are all consistent with zero. Fits to all these HERMES and COMPASS data, including the deuteron data using isospin symmetry, indicate that the Sivers function ($f_{1T}^\perp$) for $u$ quarks in a proton is negative and for a $d$ quark in a proton is positive and approximately equal in absolute value~\cite{Anselmino:2011gs}. 
This fits the expectations from the limit of a large number of colors $N_c$~\cite{Pobylitsa:2003ty,Drago:2005gz}:
\begin{equation}
f_{1T}^{\perp u}(x,\boldsymbol k_\perp^2) = - f_{1T}^{\perp d}(x,\boldsymbol k_\perp^2) + {\cal O}(1/N_c).
\end{equation}
The flavor singlet combination of $u$ and $d$ is of the same order as the gluon contribution in $N_c$ counting~\cite{Efremov:2000ar}. The latter is thus $1/N_c$ suppressed w.r.t\ the flavor non-singlet quark Sivers effect at not too small $x$ ($x\sim 1/N_c$)~\cite{Efremov:2004tp}.

Within the current accuracy, the SIDIS data do not require any sea quark or gluon contributions, which among other considerations (see section \ref{theoryconstraints}) led Brodsky and Gardner to conclude that the gluon Sivers function is small or even zero (``absence of gluon orbital angular momentum'')~\cite{Brodsky:2006ha}. The SIDIS data from HERMES, COMPASS, and Jefferson Lab Hall A are of course at rather modest $Q^2$ and not too small $x$ values, i.e.\ in the valence region. One cannot yet draw any conclusions about the gluon Sivers function at higher $Q^2$ and smaller values of $x$. Moreover, the data
certainly still allow for gluon Sivers contributions of the order of $1/N_c$ times the valence quark Sivers functions. This is evident from the fits by Anselmino {\it et al.}~\cite{Anselmino:2011gs}, where the first transverse moment of the $u$ and $d$ Sivers functions have error bands that are at least around 30\% of the central values.

Note that the SSA in the ``inclusive'' process $e \, p \to h \, X$, where the back-scattered lepton is not observed~\cite{Airapetian:2013bim,Allada:2013nsw}, does not allow for an interpretation in terms of TMDs, as the data are dominated by $Q^2 \approx 0$. Even for large $p_T$ the appropriate factorization would be collinear factorization and the Sivers type of asymmetry would probe the Qiu-Sterman functions instead~\cite{Kang:2011jw}, which as discussed above have some relation to the Sivers TMDs, but only via the tail or possibly via the first transverse moment. The asymmetries for $p_T > 1\ {\rm GeV}$ are found to be at the level of 5-10\% for positive hadrons. Fits will need to make clear how much room there is for a gluon Qiu-Sterman effect. Given the fact that the gluon Qiu-Sterman function does not enter at leading order in $\alpha_s$ in this process, this room may be considerable.

\section{Sivers asymmetry in other processes}

Several other $p\, p$ scattering processes to access the gluon Sivers function have been suggested over the past years:
$p^\uparrow\,p\to {\rm jet}\, {\rm jet}\, X$~\cite{Boer:2003tx}, $p^\uparrow\,p\to D\, X$~\cite{Anselmino:2004nk,Kang:2008ih,Beppu:2013uda},
$p^\uparrow\,p \to \gamma \, X$~\cite{Schmidt:2005gv}, $p^\uparrow\,p \to \gamma \,\rm jet\, X$~\cite{Schmidt:2005gv,Bacchetta:2007sz}, $p^\uparrow\,p \to \gamma^* \, X\to \mu^+\, \mu^-\, X$~\cite{Schmidt:2005gv},
$p^\uparrow\,p\to {\rm jet}\, X$\footnote{Single transverse spin asymmetries in jet production measured at RHIC \cite{Adamczyk:2012qj,Bland:2013pkt} at forward rapidities (the valence region) show very small asymmetries, which is probably due to a cancellation among $u$ and $d$ quark contributions \cite{Anselmino:2013rya}.}, $p^\uparrow\,p\to \pi\,{\rm jet}\, X$~\cite{D'Alesio:2010am}, $p^\uparrow\,p\to \eta_{c/b} \, X$~\cite{Schafer:2013wca}.
Several of these processes are like $A_N$ in (high-$p_T$) pion production, which means that they deal with twist-3 collinear factorization and only provide indirect or limited information about the gluon Sivers TMD.
Several other processes run into the problem of TMD factorization breaking contributions~\cite{Rogers:2010dm} and hence are not safe. In principle they do probe TMDs but as a result of TMD factorization breaking contributions, conclusions about the gluon Sivers function from their measurements cannot be drawn safely. This applies for instance to the process $p^\uparrow \, p \to {\rm jet}\, {\rm jet}\, X$ (measured at RHIC to be small at the few percent level~\cite{Abelev:2007ii}), which moreover suffers from cancellations
between $u$ and $d$ contributions and between the effects of initial and final state interactions \cite{Bomhof:2007su,Qiu:2007ey,Vogelsang:2007jk}. TMD factorization breaking would also apply to open heavy quark production: $p^\uparrow\,p \to Q\, \overline{Q} \, X$, such as $p^\uparrow\,p \to D^0_{}\, \overline{D}{}^0 \, X$, cf.\ e.g.\ \cite{Catani:2014qha}. Whether the problem also applies to double heavy quarkonium production remains to be seen, because in practice the color singlet contributions may give the dominant contribution in that case.
Among the hadronic collisions the processes having one or two color singlets in the final state would in any case be safest. One very promising example is $p^\uparrow\,p \to \gamma \,{\rm jet}\, X$~\cite{Bacchetta:2007sz}, where it depends on the rapidity of the photon and the jet, i.e.\ on the $x$ fraction of the parton in polarized proton, whether the gluon Sivers function dominates over the quark one or vice versa. Another very promising example is $p^\uparrow\,p \to J/\psi \, \gamma \, X$, which is predominantly initiated by gluon-gluon scattering which is an order in $\alpha_s$ higher than the gluon contribution in $p^\uparrow\,p \to \gamma \,{\rm jet}\, X$) and for which the color singlet contribution dominates over the color octet one to a large extent~\cite{Dunnen:2014eta,Lansberg:2014myg}. The same applies to $p^\uparrow\,p \to J/\psi \,J/\psi \, X$ (see the contribution by Lansberg and Shao in this special issue). AFTER@LHC would be very well suited for studying these processes.

SSA experiments could be done at AFTER@LHC where the beam of protons or lead ions of the LHC would collide with a fixed target that is transversely polarized. Such $p\,p^\uparrow$ and $Pb\, p^\uparrow$ collisions would have a center-of-mass energy $\sqrt{s_{NN}}$ of 115 and 72 GeV, respectively, and have high luminosity and good coverage in the rapidity region of the transversely polarized target (mid and large $x^\uparrow_p$)~\cite{Brodsky:2012vg}. Polarized Drell-Yan and prompt photon production studies could be done to measure the quark Sivers function very precisely, perhaps to the level that the gluon Sivers function becomes relevant, despite the large values of $x$ in the polarized target. As mentioned $\gamma \,{\rm jet}$ and $J/\psi\, \gamma$ production could be used to study the gluon Sivers effect directly, where the former would need specific selection of the rapidities. In addition, the comparison of $Pb\, p^\uparrow\, \to \gamma \,{\rm jet}\, X$ and $p\, p^\uparrow \to \gamma \,{\rm jet}\, X$ would give a further handle on determining the relative sizes of quark and gluon Sivers functions. Other processes, such as $D$-meson or $J/\psi$ production, would allow a similar study of Qiu-Sterman functions, including the tri-gluon ones, which are of course interesting in their own right. See \cite{Kanazawa:2015fia}
for a more detailed and quantitative study of twist-3 transverse single-spin asymmetries in proton-proton collisions at the AFTER@LHC experiment. All these possibilities offer a very interesting complementary opportunity or even a competitive alternative to the other existing high-energy particle physics spin projects aiming at studying the role of gluons in transversely polarized protons.

In electron-proton scattering one of the most promising processes to directly probe the gluon Sivers function is open charm production, $e \, p^\uparrow \to e'\, c \bar{c} \, X$, which could ideally be studied with an Electron-Ion Collider (EIC).
By selecting the charm (or bottom) quark, one effectively eliminates the subprocesses $\gamma^* q \to q g$ and $\gamma^* \bar{q} \to \bar{q} g$ and becomes essentially\footnote{This assumes that intrinsic charm contributions are suppressed by selecting sufficiently small $x$ values.} sensitive to $\gamma^* g \to c \bar{c}$, and thus to the gluon Sivers function\footnote{A similar argument is used in the study of high-$p_T$ hadron pairs in muon-deuteron and muon-proton scattering~\cite{Szabelski:2015sza,Szabelski:2015qnp}, where photon-gluon fusion is expected to dominate. The relevant asymmetry $A_{UT}^{\sin(\phi_{2h}-\phi_S)}$ is found to be $-0.14 \pm 0.15 {\rm (stat.)} \pm 0.06 {\rm (syst.)}$ at $\langle x_G \rangle = 0.13$ for the deuteron and $-0.26 \pm 0.09 {\rm (stat.)} \pm 0.08 {\rm (syst.)}$ at $\langle x_G \rangle = 0.15$ for the proton. For the interpretation of the data in terms of the gluon Sivers effect, $Q^2$ and the $p_T$ of each hadron need to be sufficiently large to trust factorization.}.
Here the transverse momenta of the heavy quarks are considered to be almost back-to-back. There is no problem with TMD factorization breaking contributions of the type discussed in~\cite{Rogers:2010dm}, but that does not mean the process is as straightfoward as SIDIS. Even in the case where one considers charm jets, one has to include a description of the transverse momentum distribution inside such a jet. It may be easier to consider $D^0_{} \overline{D}{}^0$ measurements (for a study of the twist-3 SSA in large $p_T$ $D$ meson production in SIDIS, i.e.\ $e \, p^\uparrow \to e'\, D \, X$, see~\cite{Kang:2008qh,Beppu:2012vi}). 
In either case one deals with 3 TMDs. Such processes involve a different soft factor (in this case a vacuum correlator of 6 Wilson lines) compared to processes involving 2 TMDs as in SIDIS, affecting the predictability.
This has been discussed at the one-loop level in~\cite{Zhu:2013yxa}.
The SSA in $e \, p^\uparrow \to e'\, D^0_{}\, \overline{D}{}^0 \, X$ has been studied for some models of the gluon Sivers function in~\cite{Boer:2011fh}, cf.\ section 2.3.1. This may be the `smoking gun' process for the gluonic Sivers effect at an EIC. It  should be mentioned though that it actually probes a different gluon Sivers TMD than the hadronic processes discussed above. This is discussed in the next section. It shows that hadronic processes are complementary to DIS processes.

For completeness we mention that when comparing extractions of the gluon Sivers TMD from different processes, one not only has to take care of the process dependence, but also of the different energy scales. Under TMD evolution from one scale to another, the transverse momentum distribution changes. For details we refer to~\cite{Kang:2011mr,Aybat:2011ge,Anselmino:2012aa,Sun:2013dya,Boer:2013zca,Echevarria:2014xaa}.
  
\section{Process dependence of the gluon Sivers function\label{sect:procdep}}

Once a set of processes that in principle allow to probe the gluon Sivers TMD has been obtained, one still has to take into account the fact that such a TMD is process dependent. For quarks the famous overall sign change between the Sivers TMD probed in SIDIS and the one probed in Drell-Yan is expected~\cite{Brodsky:2002cx,Brodsky:2002rv,Collins:2002kn,Belitsky:2002sm} and is currently under experimental investigation. For gluons the situation is more complicated as each gluon TMD depends on two gauge links (in the fundamental representation), so there are more possibilities~\cite{Bomhof:2006ra,Bomhof:2007xt,Buffing:2013kca}.
The gauge link structure of the gluon distributions in $e \, p \to e'\, D^0\, \overline{D}{}^0 \, X$ differs from the one in for instance
$p \, p \to \gamma \, {\rm jet}\, X$ (cf.~\cite{Dominguez:2011wm} for the comparison at small $x$). Clearly, this will complicate the analysis of gluon Sivers effect which will involve more than one gluon Sivers function. In~\cite{Buffing:2013kca} it was demonstrated that any gluon Sivers function can be expressed in terms of two ``universal'' gluon Sivers functions,
\begin{equation}
f_{1T}^{\perp g [U]}(x,\boldsymbol k_\perp^2) = \sum_{c=1}^2 C_{G,c}^{[U]} \,f_{1T}^{\perp g (Ac)}(x,\boldsymbol k_\perp^2),
\end{equation}
where the coefficients $C_{G,c}^{[U]}$ are calculable for each partonic subprocess. The first transverse moments of the two distinct gluon Sivers functions are related (at least at tree level) to the two distinct tri-gluon Qiu-Sterman functions $T_G^{(f/d)}$. Therefore, we will refer to the universal gluon Sivers functions as $f^{\perp \, g\, (f)}_{1 T}$ and $f^{\perp\,g\,(d)}_{1 T}$. The
two functions have different behavior under charge conjugation, just like $T_G^{(f)}$ is a matrix element of a $C$-even operator and
$T_G^{(d)}$ of a $C$-odd operator.

The process $e \, p^\uparrow \to e'\, D^0\, \overline{D}{}^0 \, X$ is dominated by just one partonic subprocess $\gamma g \to q \bar{q}$ and thus probes the gluon Sivers function with two future-pointing ($+$) links \cite{Pisano:2013cya}, which is $f^{\perp\,g\,(f)}_{1 T}$~\cite{Buffing:2013kca}. The process $p^\uparrow \, p \to \gamma \, {\rm jet}\, X$ probes the subprocesses $qg \to \gamma q$ and $q\bar{q} \to \gamma g$. If one selects kinematics such that one probes small $x$ values in the polarized proton, such that $qg \to \gamma q$ dominates, then this process accesses the gluon Sivers with a future and a past-pointing link, which corresponds to $f^{\perp\,g\,(d)}_{1 T}$. The theoretical expectations are different for these two cases.

\section{Theoretical constraints on Sivers function \label{theoryconstraints}}

Constraints on the unintegrated gluon Sivers TMD $f^{\perp\, g}_{1 T}(x,\boldsymbol k_\perp^2)$ from fits have to take into account that it is theoretically possible that both the quark and the gluon Sivers TMD can have nodes in $x$ and/or $ k_\perp$~\cite{Boer:2011fx,Kang:2011hk}.
The possibility of a node in $x$ is supported by the observation~\cite{Schafer:2013opa} that the splitting function for $T_G^{(f)}$ is negative at small $x$, in analogy to the $\Delta g$ case. Fits to SIDIS data (studied with a rather restrictive parameterization and in a restricted kinematic range) do not appear to require a node~\cite{Kang:2012xf}, but that does not exclude this possibility. Especially when comparing data from different kinematic regions and different processes, this option should be kept in mind. Nodes can of course have a large effect on integrals of Sivers functions, such as the first transverse moment~\eqref{transvmom} and its first Mellin moment (for parton $a$)
\begin{equation}
\langle \boldsymbol k_{\perp a}\rangle =-M(\hat{\boldsymbol S}_T\times \hat{\boldsymbol P}) \int \ud x\, f_{1T}^{\perp (1) a}(x),
\end{equation}
which is the average transverse momentum inside a transversely polarized target. The notation $\langle \boldsymbol k_{\perp a}\rangle $ comes from~\cite{Burkardt:2004ur}.
This quantity is related to the Sivers shift~\cite{Boer:2011xd}, the average transverse momentum shift orthogonal to the transverse spin direction, which is normalized to the zeroth transverse moment of the unpolarized TMD $f_1^{(0)}(x)\equiv\int\ud^2k_\perp\,f_1(x,\boldsymbol k^2_\perp)$:
\begin{equation}
\langle k^y_\perp (x)\rangle_{UT_x} = M\, \frac{ f_{1T}^{\perp(1)}(x;\mu,\zeta) }{ f_1^{(0)}(x;\mu,\zeta)}.
\end{equation}
Here only the $y$-component perpendicular to the transverse spin direction $x$ is nonzero and therefore considered.
Note that the Sivers shift depends in principle on the renormalization scale $\mu$, a rapidity variable $\zeta$, but also on the path of the gauge link (see below). Furthermore, there is the theoretical issue whether the transverse moments converge. For that reason a regularized version using Bessel moments has been suggested in~\cite{Boer:2011xd}, which for quarks has been evaluated on the lattice in~\cite{Musch:2011er}.
The lattice calculation of the Sivers shift for $u-d$ are consistent with negative $u$ and positive $d$ Sivers functions in SIDIS, which are expectations that follow from general arguments on final state interactions~\cite{Burkardt:2002ks}, from a model-dependent relation to GPDs~\cite{Burkardt:2003uw,Burkardt:2003je}, see below, and from fits~\cite{Anselmino:2005ea,Anselmino:2008sga}.

As said above, $f_{1T}^{\perp (1)q}(x)$ has a (process-dependent!) relation to the Qiu-Sterman function $T_{q,F}(x,x)$, which so far only has been established at tree level and for quarks. In addition, Burkardt has suggested a model-dependent relation between
the integrated quantity $f_{1T}^{\perp (1)}(x)$ and an integral over the GPD $E(x,\xi,\Delta^2)$ at zero skewness $\xi=0$ (and only for quarks)~\cite{Burkardt:2003uw}:
\begin{equation}
f_{1T}^{\perp (1)}(x) \propto
\int \ud^2b_\perp \,\bar{\mathcal I}(\boldsymbol b_\perp)
\,\frac{\partial}{\partial  b_y} {\cal E}(x, \boldsymbol b_\perp^2),
\end{equation}
for a nucleon polarized in the transverse $x$ direction. Here ${\cal E}(x,\boldsymbol{b}_\perp^2) \equiv \int \frac{d^2 \Delta_\perp}{(2\pi)^2} \, e^{-i\boldsymbol{b}_\perp \cdot \boldsymbol{\Delta}_\perp}\, E(x,0,-\boldsymbol{\Delta}_\perp^2)$ and $\bar{\mathcal I}(\boldsymbol b_\perp)$ is called the lensing function. This relation has been obtained in
models~\cite{Burkardt:2003uw,Burkardt:2003je,Meissner:2007rx,Gamberg:2009uk}.
It allows to relate $\langle \boldsymbol k_{\perp q}\rangle$ to the anomalous magnetic moment $\kappa_q$ associated with the quark $q$:
\begin{equation}
\int \ud x \int \ud^2{b}_\perp\, {\cal E}_q(x,\boldsymbol{b}^2_\perp) = \kappa_q,
\end{equation}
albeit {\it in a model-dependent and (due to the different integrals involved) only qualitative way.} This relation does confirm the expectations for the relative signs between the $u$ and $d$ Sivers functions, and has been used in Ref.~\cite{Bacchetta:2011gx} to fit SIDIS data for the Sivers effect with the integral constrained by the anomalous magnetic moments. Interestingly, this led to a new estimation of the quark total angular momentum which turned out to be in agreement with most common GPD extractions~\cite{Guidal:2004nd,Diehl:2004cx,Ahmad:2006gn,Goloskokov:2008ib,Diehl:2013xca}. The relation between $\langle \boldsymbol k_{\perp q}\rangle$ and  $\kappa_q$ is also at the heart of the argument by Brodsky and Gardner of why a gluon Sivers function is expected to be small. Using that $\kappa_u^p = 2\kappa_p + \kappa_n = 1.673$ and $\kappa_d^p = 2\kappa_n + \kappa_p = -2.033$, one sees the opposite signs reflected, but since $|(\kappa_p+\kappa_n)/2|=0.06 \ll \kappa_{p/n} \approx 1.8-1.9$ this suggests that there is little room for gluon contributions~\cite{Brodsky:2006ha}.
If $(\kappa_u^p + \kappa_d^p)/2$ is taken as a measure for $\kappa_g^p$, the latter is about 10\% smaller than $\kappa_q^p$.
This would suggest that $\kappa_g^p$ is of order $1/N_c^2$ rather than $1/N_c$, which in turn would suggest a similar additional $1/N_c$ suppression for the gluon Sivers function. Clearly there are various (strong!) assumptions going into this type of argument, such that the conclusion can certainly not be taken at face value. Apart from the assumptions on the relation to the gluon Sivers function, it is not clear that one can use very low-energy quantities to deduce something about the size of the gluon contributions at energies around or above 1 GeV to begin with.

Burkardt derived a further constraint on the fully integrated quantity $\langle \boldsymbol k_{\perp a}\rangle$, nowadays referred to as the Burkardt sum rule (BSR)~\cite{Burkardt:2004ur}, stating that the total transverse momentum of all partons in a transversely polarized proton must vanish~\cite{Lorce:2015lna},
\begin{equation}
\langle \boldsymbol k_\perp \rangle = \sum_{a = q, \bar q, g} \langle \boldsymbol k_{\perp a} \rangle  = \boldsymbol 0\, .
\end{equation}
In terms of the Sivers function, the BSR takes the form~\cite{Efremov:2004tp}
\begin{equation}
 \sum_{a = q, \bar q, g}\int  d x \,
 f^{\perp\,(1)\, a}_{1 T} (x) = 0.
\label{eq:BSR}
\end{equation}
Its validity has been checked  explicitly in a diquark spectator model in Ref.~\cite{Goeke:2006ef}.
The fits to SIDIS data from Ref.~\cite{Anselmino:2008sga} at the scale $Q^2 = 2.4$ $\gev^2$ almost saturate the BSR already
with the $u$ and $d$ quark contributions alone:
\begin{equation}
\langle k_{\perp u} \rangle = 96^{+60}_{-28} \ {\rm MeV}, \quad
\langle k_{\perp d} \rangle \, = -113^{+45}_{-51} \ {\rm MeV}.
\end{equation}
The contributions of the sea quarks are all small and together allow the following range for the gluon contribution:
\begin{equation}
-10 \le \langle k_{\perp g} \rangle \le 48~\mev.
\end{equation}
This means there is certainly still room for a 30\% contribution from gluons w.r.t.\ the valence quarks.
Of course, it should be emphasized that these values were obtained under assumptions on the $k_\perp$ dependence, the absence of nodes, and extrapolations outside the kinematic region accessed by the SIDIS experiments.

The derivation of the BSR by Burkardt~\cite{Burkardt:2004ur} considers gauge links as appear in SIDIS and involves a gluon correlator containing
the antisymmetric $f_{abc}$ structure constant of $SU(3)$.  As shown in~\cite{Bomhof:2006ra,Bomhof:2007xt}, there is also a gluon correlator with the symmetric $d_{abc}$ structure constant. As a consequence, inclusion of gauge links in the operator definition of TMD distributions gives rise to two distinct gluon Sivers functions, $f^{\perp\,g\,(f)}_{1 T}$ and $f^{\perp\,g\,(d)}_{1 T}$ (corresponding to the $(Ac)$ label used in~\cite{Buffing:2013kca}). However, the BSR essentially expresses transverse momentum conservation. Since the momentum operator in QCD is $C$-even, only the gluon Sivers function $f^{\perp\,g\,(f)}_{1T}$ which is associated with a $C$-even operator is constrained by the BSR. The gluon Sivers function $f^{\perp\,g\,(d)}_{1T}$, which is associated with a $C$-odd operator, is not expected to satisfy a BSR where quark and gluon contributions cancel each other. Judging from the small $x$ behavior of the $T_G^{(d)}(x,x)$ expected from its evolution equation, the integral of $f^{\perp\,g\,(d)}_{1 T}$ over $x$ may even not converge.

\section{Conclusions}

In summary, no hard constraints on the size of the gluon Sivers function exist apart from the positivity bound, although the theoretical expectation from large $N_c$ considerations (expected to hold approximately for not too small $x$) favors a 30\% gluon to quark Sivers ratio, which is still completely allowed by all SIDIS data, including the COMPASS deuteron data. It may turn out that the ratio is smaller, but much smaller than 10\% may in turn be considered unnaturally small. Strictly speaking, no direct conclusion about the size of the gluon Sivers function can be drawn from $A_N$ data. Like $p^\uparrow \, p \to \pi \, X$, many other processes suggested in the literature to probe the gluon Sivers function actually deal with collinear factorization and as such they are sensitive to complicated linear combinations of quark-gluon and tri-gluon Qiu-Sterman functions (and chiral-odd and fragmentation function versions of them) rather than to Sivers functions directly. Inferring constraints on the gluon Sivers function, even on its large transverse momentum tail must therefore be done with much care. In the Generalized Parton Model description of $A_N$ at midrapidity, the {\it effective} gluon Sivers function is currently still allowed to be 30\% of the up quark Sivers function, despite the smallness of the asymmetry. Other suggested processes that in principle probe TMDs may suffer from TMD factorization breaking contributions and any results on the gluon Sivers function from measurements of such processes cannot be trusted. This applies for instance to the process $p^\uparrow \, p \to {\rm jet}\, {\rm jet}\, X$.

The most promising processes that directly give access to the gluon Sivers effect are $p^\uparrow\,p \to \gamma \,{\rm jet}\, X$, $p^\uparrow\,p \to J/\psi \, \gamma X$ and $e \, p^\uparrow \to e'\, c \, \bar{c} \, X$. The first process can be studied at RHIC and at a polarized fixed-target experiment at LHC (AFTER@LHC), the second process also at AFTER@LHC, and the third process at a possible future Electron-Ion Collider. Due to the inherent process dependence of TMDs, the gluon Sivers TMD probed is in principle different in these processes. They can be expressed in terms of two universal gluon Sivers functions that appear in different linear combinations in different processes. Extracting and comparing these universal functions is very interesting from a theoretical point of view. The fact that a difference can exist is a consequence of the non-Abelian nature of QCD. Both functions satisfy different theoretical constraints. Although TMD factorization is expected to hold for these processes, that has not been demonstrated yet to all orders. Apart from the process dependence, there is also the issue of modified soft factors to contend with still. Nevertheless, as far as experimentally demonstrating and measuring a gluon Sivers effect in transversely polarized protons, several complementary future possibilities exist, in which AFTER@LHC can play  a very important role.

\section*{Acknowledgments}
We wish to thank Mauro Anselmino, Maarten Buffing, Umberto D'Alesio, Jean-Philippe Lansberg, Piet Mulders, and Francesco Murgia for useful discussions and/or feedback on the text. C.L. acknowledges support by the Belgian Fund F.R.S.-FNRS \emph{via} the contract of Charg\'e de Recherches. C.P.\ acknowledges support by the ``Fonds Wetenschappelijk Onderzoek - Vlaanderen'' (FWO) through a postdoctoral Pegasus Marie Curie Fellowship.

\printindex

\end{document}